# Magnetic, Structural and cation distribution studies on FeO·Fe$_{(2-x)}$Nd$_x$O$_3$ (x=0.00, 0.02, 0.04, 0.06 and 0.1) nanoparticles.

W. W. R. Araujo[*], C. L. P. Oliveira, G. E. S. Brito, A. M. Figueiredo Neto

Institute of Physics, University of São Paulo, Rua do Matão 1371, 05508-090, São Paulo, SP, Brazil.

J. F. D. F. Araujo

Department of Physics, Pontifical Catholic University of Rio de Janeiro, RJ, Brazil.

**Abstract**

We synthesized and characterized the colloidal suspensions of FeO·Fe$_{(2-x)}$Nd$_x$O$_3$ nanoparticles with x=0.00, 0.02, 0.04, 0.06 and 0.1. The effect of the Fe$^{3+}$ ion replacement by Nd$^{3+}$ on the crystal structure is in-depth studied, through X-ray diffraction (XRD) and the obtained cation distribution. The magnetic properties of the synthesized FeO·Fe$_{(2-x)}$Nd$_x$O$_3$ nanoparticles also were investigated and corroborated by other physical methods. A remarkable saturation magnetization of 105 Am$^2$/kg was achieved for x=0.06.

**I. Introduction**

There has been an enormous research effort recently in colloidal magnetic nanocrystals, so that hundreds of papers are published per year in this field of scientific research [1]. These nanocrystals can be obtained with controlled size, shape and composition [2, 3]. Colloidal magnetic nanocrystals (CMNs) attract increasing interest both in fundamental sciences and in technological applications [4]. Colloidal magnetic nanocrystals (CMNs) are suspensions of magnetic nanoparticles (MNPs) dispersed in a liquid carrier [4, 5]. When the particle dimension is small (typically ~10 nm), the particle presents a single magnetic domain, with a large magnetic moment called superspin [6-8]. The surface of MNPs can be modified by several stabilizing agents or functional groups. Once, the surface of the magnetic nanoparticles is modified, they become highly functional materials. Some of the MNPs physical properties can be controlled by external magnetic fields or magnetic-field gradients [9, 10], that is, they are stimuli-responsive systems. Some applications of ferrofluids are separation media [1], heat-conduction media [4], gas fluidized beds [1], sealants [2, 7, 11] and hydraulic car suspensions [7]. MNPs can also be used for drug delivery, medical diagnosis, and cell destruction [4]. One of the peculiarities of a ferrofluid is how its nanoscopic organization is affected by an applied external magnetic field [10]. In the field of magneto-optical devices some applications can be found, for integrated optics [9, 11], optical fibers [9, 11] and tunable beam splitter [11].

Magnetic nanoparticles are free to move when dispersed in a liquid carrier medium, and different physical event may occur as a function of an applied magnetic field [12]. The blocking temperature ($T_B$) defines when the system of MNPs passes from blocked to superparamagnetic (SPM) state [12, 13]. The Zero-field cooled (ZFC) and Field cooled (FC) technique is widely used to study granular materials. ZFC-FC curves allows the determination of the blocking temperature of the system $T_B$ [12, 13]. The Small Angle X-ray Scattering is used to study structural parameters of fluid samples from the analysis of the experimental scattering pattern [2, 9]. For polydisperse systems, the experimental scattering pattern corresponds to an average that is performed over the particles in the solution. Thus, the experimentally reached values correspond to an average over the entire ensemble of particles rather than a single particle [2, 9]. To tackle this problem, ZFC-FC curves complemented by SAXS scattered intensities could supply several insights about the MNPs investigated.

Rare-Earth doping significantly alters the nucleation and growth of nanoferrites, which facilitate magnetic spin orientation [2, 14]. It is known that the magnetic behavior of spinel ferrite compounds is mostly due to the interaction between the iron atoms [14]. Rare-Earth ions are more favorable to enter in octahedral sites of the spinel structure; this causes 4f-3d interactions that promotes structural distortion, lattice strain and changes in saturation magnetization [14]. The literature reports the use of Neodymium (Nd) for doping copper nanoferrites [14]. Mixed manganese−neodymium−copper (Mn–Nd–Cu) nanoferrites was produced by sonochemical method [14]. Aslam and co-workers performed a co-doping of $Nd^{3+}$ and $Pr^{3+}$ on lithium nanoferrite and reported the effects on the magnetic and structural properties of the system [15]. The effect of $Nd^{+3}$ doping on Mn-Zn ferrite was reported in the literature [16, 17], and an enhancement of the saturation magnetization, due to the new cation distribution imposed by Nd doping was reported. Jain and co-workers studied the influence of rare earth ions on structural, magnetic and optical properties of magnetite nanoparticles [18]. Jain also reported that there is a variation in the saturation magnetization maximum value, directly proportional to the number of unpaired 4f electrons in the dopant element [18]. Huan and co-workers found similar results for $RE^{3+}$-doped $Fe_3O_4$ samples ($RE^{3+}=Ln^{3+}$, $Eu^{3+}$ and $Dy^{3+}$) [19].

In the present study we report on the synthesis and characterization of colloidal suspensions of $FeO·Fe_{(2-x)}Nd_xO_3$ nanoparticles with x=0.00, 0.02, 0.04, 0.06 and 0.1. The samples were synthesized by co-precipitation method. The characterization was done by: X-ray diffraction (XRD), Transmission Electron Microscopy (TEM), Small Angle X-ray Scattering (SAXS), Optical bandgap and Zero-field and Field Cooling magnetization (ZFC-FC). Therefore, the effect of the $Fe^{3+}$ ion replacement by $Nd^{3+}$ on the crystal structure is in-depth studied, through X-ray diffraction (XRD) and the obtained cation distribution. The magnetic properties of the synthesized $FeO \cdot Fe_{(2-x)}Nd_xO_3$ nanoparticles also were investigated and corroborated by other physical methods.

## II. Material and methods

### A. Materials reagents

The materials used to obtain the MNPs were: $FeCl_3 \cdot 6H_2O$, ≥99%; $FeCl_2 \cdot 4H_2O$, ≥99%; $NdCl_3 \cdot 6H_2O$, ≥99%; *cis*-9-Octadecenoic acid (Oleic acid), ≥99%; NaOH, ≥99%; and kerosene. All primary materials were acquired from Sigma-Aldrich and used as received.

### B. Synthesis procedure by co-precipitation method

The magnetic nanoparticles were synthesized by co-precipitation of an aqueous mixture of $FeCl_3$ (ferric chloride) and $FeCl_2$ (ferrous chloride) salts and stabilized at pH~12 [11, 20-24]. Briefly, ferric chloride (4.00 mmol) and ferrous chloride (2.00 mmol) in a 2:1 molar ratio and 8 mL of oleic acid are mixed in 25 mL of deionized water. The solution was heated up to 80 °C and magnetically stirred for 30 min. Then, 30 mL of NaOH was added to the solution to precipitate the particles at room conditions, under vigorously stirred for more 30 min. When the pH reaches ~ 11, a severe reaction occurs and the solution becomes dark brownish. Thereafter, the resultant solution was cooled to room temperature. At last, the MNPs were precipitated with a permanent magnet and then washed ten times with deionized water to remove residual unreacted salts. The procedure described above was used to produce magnetite nanoparticles with x = 0.00 and that samples was called as FF-REF.

The samples FF-ND1, FF-ND2, FF-ND3 and FF-ND5 are labeled according to molar percentage amount of substitution of $Fe^{3+}$ ions by $Nd^{3+}$ ions. These percentages are 1, 2, 3 and 5% (x=0.02, 0.04, 0.06 and 0.10), respectively. FF-ND1, FF-ND2, FF-ND3 and FF-ND5 were prepared with an aqueous solution composed of $FeCl_3 \cdot 6H_2O$ ((4 - i) mmol), $NdCl_3 \cdot 6H_2O$ (i=(0.04, 0.08, 0.12 and 0.20) mmol), $FeCl_2 \cdot 4H_2O$ (2.00 mmol) and 8 mL cis-9-octadecene. Then, 30 mL of NaOH was added to the solution to precipitate the particles at room conditions, under vigorously stirring for 30 min. The remaining procedures were the same used to prepare the FF-REF sample.

Hereafter, the doped concentration (x=0.02, 0.04, 0.06 and 0.10) of $Nd^{3+}$ in magnetite (x=0.00, $Fe_3O_4$) will be denoted as FF-ND1, FF-ND2, FF-ND3 and FF-ND5, respectively. All samples were sterically stabilized with a single oleic acid layer, chemisorbed on the particles' surfaces. MNPs are dispersed in kerosene.

### C. X-ray Diffraction

Powder X-ray Diffraction (XRD) patterns were obtained to investigate nanoparticles' crystalline structure. XRD was done in a Brucker-AXS D8 series 2 diffractometer, set to a Bragg Brentano Parafocussing Geometry. A Cu Ka source ($\lambda$ = 1.5414 Å) generated X-rays at room temperature. The difractometer was operated at 40 kV, 30 mA. The experimental pattern data were registered in continuous scan mode, scattering angle 2θ from 15 to 80º, in steps of 0.02º.

**D. Transmission Electron Microscopy**

The overall form (morphology) and size distribution of the MNPs were examined by transmission electron microscopy (TEM). TEM was performed on a JEOL 1010 (Japan) microscope with an acceleration voltage of 80 kV. TEM micrographs were acquired by a Gatan Bioscan 782 CDD camera of 1K x 1K pixels. The colloidal suspension of MNPs were prepared by diluting the original colloidal suspensions 100 times and maintained in an ultrasonic bath for 30 min. A drop of the colloidal suspension was placed on a Formvar™ coated 200 mesh copper grid. The residual excessed fluid was blotted and dried until the time that the grids was examined into microscope. MNPs number-weighted-size distributions were obtained by measuring about ~500 MNPs with the ImageJ freeware [25]. The number-weighted size-distribution data were fitted to a log-normal distribution function given by [2]:

$$p(D)dD = \frac{1}{D \cdot s\sqrt{2\pi}} exp\left(\frac{-ln^2\left(\frac{D}{D_m}\right)}{2 \cdot s^2}\right) \quad (1)$$

where $D_m$ and $s$ are fitting parameters. The number-weighted mean diameter, $\langle D_N \rangle$ and standard deviation of particle size, σ, are written as [2]:

$$\langle D_N \rangle = D_m exp\left(\frac{s^2}{2}\right) \quad (2)$$

$$\sigma_N = \langle D_N \rangle (\sqrt{exp(s^2) - 1}) \quad (3)$$

**E. Small Angle X-ray Scattering**

The Small Angle X-ray Scattering (SAXS) patterns of the solutions were investigated. The setup used to data acquisition was a Xeuss (Xenocs™). The Xeuss is equipped with a microfocus x-ray source Genix, with radiation of $\lambda$ = 1.5414 Å (Cu). The system uses two scatterless slits for beam collimation and it reaches the sample with a square cross section of 0.4 x 0.4 mm². The primary and scattered beams remain in a vacuum ($10^{-2}$ mbar) chamber to avoid scattering by the air. Each sample were mounted in a cylindrical Mark-tubes of quartz glass capillary (Hilgenberg, 1.5 mm outer diameter) with principal axis in the vertical direction. The measured 2D scattered data were recorded by a Pilatus (Dectris) 300 K 20 Hz 2D detector. The exposition time was 600 s and all measurements were performed at room temperature (~ 22 °C).

The 1D scattering intensity versus scattering vector module, $I(q)$ and $q$, defined by $q = (4\pi \sin\theta)/\lambda$ was obtained by averaging the data over a 20° slice in horizontal and vertical directions. The data treatment, blank subtraction, and data normalization were performed with the software SUPERSAXS (C.L.P. Oliveira and J. S. Pedersen, unpublished). The contribution of blank (kerosene) was independently measured and subtracted from the sample data. To obtain the scattered intensity data in absolute-scale units, $cm^{-1}$, water was used as standard. The sample-detector distance was 839 cm, which allows measurements with $q$ in the range $0.01 < q < 0.35$ Å$^{-1}$ [2, 9, 12].

**F. Zero-field cooling and field-cooling (ZFC-FC)**

The magnetic properties of the MNPs were measured in a home-made Hall-effect magnetometer [26, 27]. The magnetization curves were performed at room temperature and low temperatures (6K) and under different applied magnetic fields from −2.0 to +2.0 T. The DC magnetization as a function of temperature was measured in both zero-field cooled (ZFC) and field cooled (FC) regimes. ZFC-FC protocol was performed in a temperature range from 5 to 300 K, applying a 5.0 mT magnetic field [26, 27].

**III. Results and Discussion**

**A. X-ray Diffraction (XRD)**

XRD patterns of the samples investigated are shown in FIG. 1. These results reveal a single-phase cubic spinel structure corresponding to the Fd3m space group [28]. The XRD pattern show diffraction peaks corresponding to the (220), (311), (400), (422), (511), (440), (620) and (533) crystallographic planes. These results agree with the XRD pattern of the $Fe_3O_4$.

The instrumental broadening ($\beta_{hkl}$) was corrected, using Warren's relation $\beta_{hkl}^2 = \beta_{measured}^2 - \beta_{instrumental}^2$ [29]. The average crystallite size (CS) of all the samples has been estimated using full width at half maximum (FWHM) of each diffraction peak and the Debye-Scherrer formula $CS = \dfrac{0.9\lambda}{\beta_{hkl}\cos(\theta)}$ [28-31], where $\lambda$ is the X-ray wavelength (Cu K$\alpha$, $\lambda$ = 1.5414 Å), $2\theta$ is the Bragg angle and $\beta_{hkl}$ is the FWHM of the diffraction peaks. The lattice parameter was investigated using the equation $a = d_{hkl} \cdot \sqrt{h^2 + k^2 + l^2}$ [28, 29]. The observed XRD patterns of all the samples were analyzed by Rietveld method [32], using the MAUD 2.80 software [33] to get the refinement parameters [28-32]. The peak shape was fitted with a pseudo-Voigt (pV) function in the refinement procedure [30]. The cation distribution was evaluated by refining the changes in the diffraction intensities, while locating the cation in an appropriate position. The background of each pattern was fitted by a polynomial function of order 5 [30]. The densities of the samples were estimated by XRD, using the relation $\rho_{XRD} = \dfrac{8\,MW}{a^3 N_a}$, where 8 refers to the number of atoms per unit cell of the spinel structure, MW is the molecular weight of the sample. The quantities $N_a$ and "a" are the Avogadro's number and the lattice parameter of the sample, respectively. The fitting parameters, as well as the lattice parameter ($a_{exp}$), unit cell volume V (Å$^3$), crystallite size (CS) and evaluated density $\rho_{XRD}$ (g/cm$^3$) are reported in Table I.

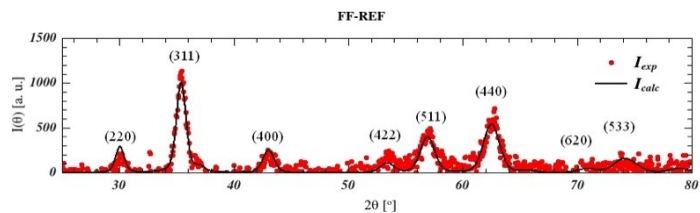

(a)

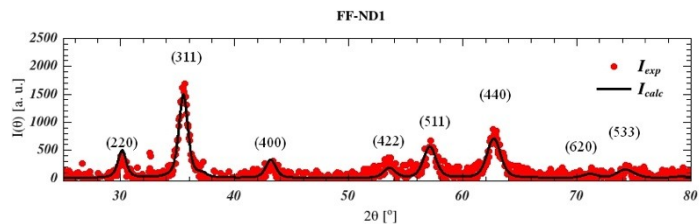

(b)

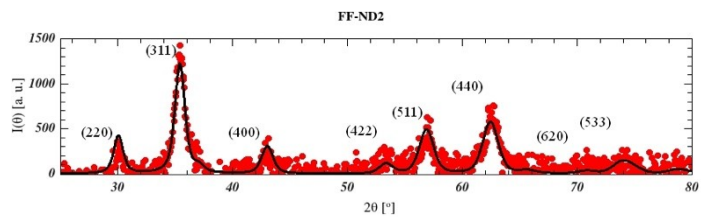

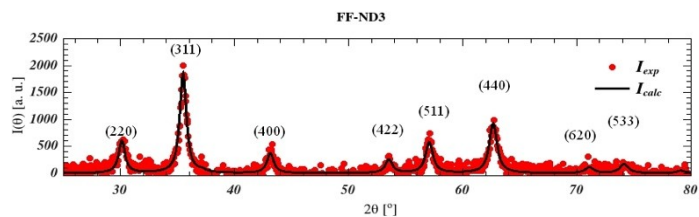

(d)

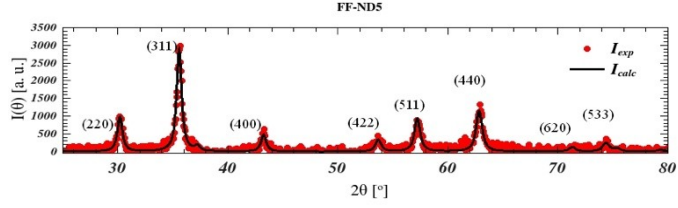

(e)

FIG. 1. XRD patterns of the magnetic nanoparticles investigated. (a) x=0.00, (b) x=0.02, (c) x=0.04, (d) x=0.06 and (e) x=0.10. Solid lines are best fits of a pseudo-Voigt (pV) function [33].

In the XRD pattern of doped samples (FF-ND1, FF-ND2, FF-ND3 and FF-ND5) we did not observe any peaks corresponding to $NdFeO_3$. The $ReFeO_3$ ($RE^{3+}$=Rare-Earth) peak was observed elsewhere due to rare earth doped ferrite [29, 30].

**Table I**

Average lattice parameter ($a_{exp}$), unit cell volume V ($Å^3$), crystallite size (CS), evaluated density $\rho_{XRD}$ (g/cm$^3$) and results of Rietveld analysis for $FeO·Fe_{(2-x)}Nd_xO_3$ nanoparticles.

| Comp. (x) | Sample | $a_{exp}$ (Å) | V (Å$^3$) | CS (nm) | $\rho_{XRD}$ (g/cm$^3$) | Rwp(%) | Rexp(%) | GoF |
|---|---|---|---|---|---|---|---|---|
| 0.00 | FF-REF | 8.39 (0.09) | 590.59 (1.31) | 23.9 (8.7) | 5.18 (0.01) | 2.23 | 1.29 | 1.73 |
| 0.02 | FF-ND1 | 8.41 (0.02) | 594.82 (0.29) | 18.3 (2.7) | 5.24 (0.01) | 2.26 | 1.24 | 1.81 |
| 0.04 | FF-ND2 | 8.42 (0.03) | 596.94 (0.44) | 15.6 (2.2) | 5.27 (0.01) | 2.60 | 1.26 | 2.06 |
| 0.06 | FF-ND3 | 8.44 (0.03) | 601.21 (0.44) | 16.2 (4.6) | 5.39 (0.01) | 1.58 | 1.32 | 1.19 |
| 0.10 | FF-ND5 | 8.46 (0.03) | 605.49 (0.44) | 14.9 (6.4) | 5.56 (0.01) | 1.83 | 1.23 | 1.48 |

The ionic radius of $Nd^{3+}$ is 0.098 nm while the ionic radius of $Fe^{3+}$ is 0.067 nm, which is about 1.46 times smaller than the dopant radius. Therefore, replacement of $Fe^{3+}$ by larger $Nd^{3+}$ ions causes an expansion of unit cell [30]. This causes an increase of the lattice parameter ($a_{exp}$), as shown in Table I, for FF-ND samples as compared with FF-REF sample. The robustness of fit (GoF), the weighted profile factor ($R_{wp}$) and expected weighted profile factor ($R_{exp}$), assure the reliability of the fits, since, low values of GoF were obtained [34,35].

## B. XRD Cation Distribution

Several physical properties of a crystal can be accessed through the knowledge of the cation distribution. Experimental Techniques such as: X-ray diffraction (XRD) pattern [30, 36-38], X-ray magnetic circular dichroism (XMCD) [39] and X-ray absorption spectroscopy [39] can be used to estimate the cation distribution for spinel ferrite materials. The cation distribution in the present work, was obtained from X-ray diffraction pattern analysis. Experimental intensity ratios were compared with the calculated intensity ratios, according to Bertaut et al.[40] method. In this method, pairs of reflections are selected according to the expression [40].

$$\frac{I_{hkl}^{exp.}}{I_{h'k'l'}^{exp.}} = \frac{I_{hkl}^{calc.}}{I_{h'k'l'}^{calc.}} \tag{4}$$

where $I_{hkl}^{exp.}$ and $I_{hkl}^{Calc.}$ are the experimental and calculated intensities for reflections (hkl), respectively. We used the intensity ratios corresponding to the planes (220), (400), (440), which are known to be sensitive to the cation distribution [30, 36-38]. In order to obtain the best-simulated/evaluated structure, the R factor was defined, according to Eq. (5).

$$R = \left| \left( \frac{I_{hkl}^{exp.}}{I_{h'k'l'}^{exp.}} \right) - \left( \frac{I_{hkl}^{calc.}}{I_{h'k'l'}^{calc.}} \right) \right| \tag{5}$$

The determination of the structure is attained by varying the cation distribution in the calculated intensity in such a way that the R factor will be minimized [30, 36-38].

The relative integrated intensity of the XRD lines can be calculated using Eq. (6):

$$I_{hkl} = L_p(\theta) \cdot |F_{hkl}|^2 \cdot P \tag{6}$$

where $I_{hkl}$ corresponds to the relative integrated intensity. The quantity $F_{hkl}$ is the structure factor, while P is the multiplicity factor for the plane (hkl), and $L_p$ is a Lorentz polarization factor (Eq. (7)), and it will be a function of the Bragg diffraction angle. The multiplicity factor was obtained from the literature [41]:

$$L_p(\theta) = \frac{1 + cos^2(\theta)}{sin^2(\theta)cos(\theta)} \tag{7}$$

The structure factor of the spinel ferrite has 24 divalent and trivalent cations and 32 oxygen anions [45]. The structural factors were calculated by using the equation proposed by Furuhashi et al.[46], Eq. (8):

$$|F_{hkl}|^2 = A_{hkl}^2 + B_{hkl}^2 \tag{8}$$

where $A_{hkl}^2$ and $B_{hkl}^2$ are related to crystal planes $hkl$ and can be determined with Eq. (9) and (10):

$$A_{hkl} = \sum_i f_i cos(2\pi(hx_i + ky_i + lz_i)) \qquad (9)$$

and

$$B_{hkl} = \sum_i f_i cos(2\pi(hx_i + ky_i + lz_i)) \qquad (10)$$

For evaluation of the atomic scattering factor we used values reported in the International Tables for X-ray Crystallography [42]. The temperature and absorption factors were neglected in our evaluation because at room temperature these factors do not affect the relative XRD intensity calculations [36]. In general, spinel structures have a high melting temperature. So, small thermo-vibrational effect of spinel on XRD patterns is expected [43].

In the present evaluation all possible cation configurations were considered with 0.01 stoichiometric sensitivities that $Nd^{3+}$ and $Fe^{3+}$ ions can site in both tetrahedral and octahedral sites, according with Eq. (11) [30]:

$$(Fe^{3+}_{(1-\delta)}Nd^{3+}_{\delta})^A[\ Fe^{2+}Fe^{3+}_{(1-\gamma)}Nd^{3+}_{\gamma}]^B O_4 \qquad (11)$$

where $x = \gamma + \delta$ and $x$ are the molar stoichiometric amount of replacement of $Fe^{3+}$ ions by $Nd^{3+}$. The closest correspondence with the actual sample structure was achieved by varying the cation distribution of the calculated intensity, which will provide a minimum R factor. (Eq. (5)) [30]. The cation distribution, the corresponding relative intensities of experimental and calculated XRD lines are given in Table II.

### Table II
Cation distribution of FeO·Fe$_{(2-x)}$Nd$_x$O$_3$ nanoparticles, experimental and calculated ratios between peak intensities.

| Comp. (x) | A-site | B-site | $I_{220}/I_{440}$ | | $I_{220}/I_{400}$ | |
|---|---|---|---|---|---|---|
| | | | Exp. | Calc | Exp. | Calc |
| 0.00 | $(Fe^{3+}_{1.00})$ | $[Fe^{2+}_{1.00}Fe^{3+}_{1.00}]$ | 0.54 | 0.52 | 1.15 | 1.03 |
| 0.02 | $(Fe^{3+}_{1.00})$ | $[Fe^{2+}_{1.00}Fe^{3+}_{0.98}Nd^{3+}_{0.02}]$ | 0.70 | 0.66 | 1.50 | 1.33 |
| 0.04 | $(Fe^{3+}_{1.00})$ | $[Fe^{2+}_{1.00}Fe^{3+}_{0.96}Nd^{3+}_{0.04}]$ | 0.73 | 0.71 | 1.38 | 1.29 |
| 0.06 | $(Fe^{3+}_{0.95}Nd^{3+}_{0.05})$ | $[Fe^{2+}_{1.00}Fe^{3+}_{0.99}Nd^{3+}_{0.01}]$ | 0.65 | 0.61 | 1.62 | 1.57 |
| 0.10 | $(Fe^{3+}_{0.94}Nd^{3+}_{0.06})$ | $[Fe^{2+}_{1.00}Fe^{3+}_{0.915}Nd^{3+}_{0.05}]$ | 0.89 | 0.83 | 2.04 | 2.03 |

The mean ionic radii in the tetrahedral (rA) and octahedral (rB) sites were calculated by using Eqs. (S1) and (S2) from Electronic Supplementary Information (ESI) file. The values of the ionic radius for $r_{Nd3+}, r_{Fe3+}$ and $r_{Fe3+}$ were taken from the literature [44], 0.98 Å, 0.67 Å, and 0.49 Å, respectively. The value of the oxygen positional parameter $u$ can be determined with Eq. (S3)(ESI), where $R_O$ is the radius of oxygen ion (1.32 Å) [25,31,39]. The theoretical lattice constant ($a_{th}$) is calculated by using Eq. (S4)(ESI) [30,36,45].

The structural parameters: tetrahedral bond length ($d_{AL}$); octahedral bond length ($d_{BL}$); tetrahedral edge length ($d_{AE}$); shared ($d_{BE}$) and unshared ($d_{BEU}$) octahedral edge length; the tetrahedral and octahedral jump length ($L_A$ and $L_B$) were calculated using Eqs. (S5)-(S11)(ESI). The results are given in Table III.

**Table III**
Theoretical parameters based on the proposed cation distribution. See text for the symbols.

| Comp (x) | rA (Å) | rB (Å) | $a_{th}$ (Å) | u (Å) | $d_{AL}$ (Å) | $d_{BL}$ (Å) | $d_{AE}$ (Å) | $d_{BE}$ (Å) | $d_{BEU}$ (Å) | $L_A$ (Å) | $L_B$ (Å) |
|---|---|---|---|---|---|---|---|---|---|---|---|
| 0.00 | 0.630 | 0.560 | 8.400 | 0.384 | 1.950 | 2.022 | 3.184 | 2.745 | 2.969 | 3.631 | 2.965 |
| 0.02 | 0.630 | 0.564 | 8.409 | 0.384 | 1.950 | 2.030 | 3.184 | 2.761 | 2.976 | 3.641 | 2.973 |
| 0.04 | 0.630 | 0.567 | 8.418 | 0.384 | 1.950 | 2.034 | 3.184 | 2.769 | 2.980 | 3.646 | 2.977 |
| 0.06 | 0.648 | 0.562 | 8.432 | 0.385 | 1.968 | 2.031 | 3.213 | 2.752 | 2.987 | 3.653 | 2.983 |
| 0.10 | 0.651 | 0.573 | 8.468 | 0.384 | 1.971 | 2.039 | 3.219 | 2.767 | 2.997 | 3.665 | 2.993 |

Table III shows that the theoretical lattice parameter ($a_{th}$), tetrahedral bond length ($d_{AL}$), octahedral bond length ($d_{BL}$), tetrahedral edge ($d_{AE}$), octahedral edge ($d_{BE}$), unshared octahedral edge ($d_{BEU}$) increase with an increase in Nd-content. Table III also shows that tetrahedral radius (rA) and octahedral radius (rB) increase as the Nd-content increases, while the oxygen parameter $u$ remains unchanged. The anion in the spinel structure, $O^{2-}$ ions, in this case, are not in general located at a fixed position of the FCC sublattice. The anion is allowed to translate and this translation is measured by a quantity named oxygen positional parameter or anion parameter. If we assume the center of symmetry at (3/8, 3/8, 3/8) position that correspond to origin at A-site, the value of u ideal is expected to be 0.375. Therefore, changes in the value of u can be interpreted as a relaxation of the structure to accommodate the cations of different radius in the A and B sites [30, 47]. The jump (hopping) lengths, $L_A$ and $L_B$ between the magnetic ions at A-site and B-site respectively, were calculated (Table III). Since the $Fe^{3+}$ radius (0.65Å) is smaller than that of the $Nd^{3+}$ (0.98 Å), the replacement of $Fe^{3+}$ leads to an increase in rA and rB. Moreover, $L_A$ and $L_B$ increases with an increase in Nd-content. The results showed that $Nd^{3+}$ ions are present in both sites at different concentrations, with the displacement of $Fe^{3+}$ ions. Therefore, it was possible to observe changes in structural parameters like bond lengths; shared and unshared edges, among others.

The magnetic properties of the particles depend on the exchange interactions between metal ions. The bond angle and inter-ionic bond length between metal ions are the more important in the overall magnitude of the magnetic exchange interaction. The magnitude of the magnetic exchange interactions is proportional to the bond angles and inversely proportional to the inter-ionic bond lengths. The inter-ionic bond lengths i.e., cation-cation distances and cation-anion distances (FIG. S1(ESI)) were calculated using equations Eqs. (S12)-(S25)(ESI). The values obtained for inter-ionic bond lengths were used for the evaluation of bond angles between the metal ions using equation Eqs. (S21)-(S25). The values for bond angles are given in Table IV.

### Table IV
Theoretical bond angles between metal ions based on the cation distribution. All angles are given in degrees (º).

| Comp (x) | $\theta_1$ | $\theta_2$ | $\theta_3$ | $\theta_4$ | $\theta_5$ |
|---|---|---|---|---|---|
| 0.00 | 122.14 | 138.45 | 94.33 | 126.06 | 70.91 |
| 0.02 | 122.28 | 138.63 | 94.14 | 126.03 | 71.10 |
| 0.04 | 122.35 | 138.72 | 94.05 | 126.02 | 71.19 |
| 0.06 | 121.99 | 138.28 | 94.52 | 126.08 | 70.74 |
| 0.10 | 122.10 | 138.41 | 94.38 | 126.06 | 70.87 |

The bond angles $\theta_1$, $\theta_2$, and $\theta_5$ are associated with the A-B and A-A exchange interactions, while $\theta_3$, and $\theta_4$ are associated with the B-B exchange interactions (FIG. (S1)(ESI)). The observed increase in $\theta_1$, $\theta_2$, and $\theta_5$, corresponding to x=0.00 to x=0.04 (Table IV) suggests the strengthening of the A-B and A-A interactions, while $\theta_3$, and $\theta_4$ decrease indicates a weakening of the B-B interaction  Comparing the systems with x=0.06 and x=0.00, the B-B exchange interactions are enhanced ($\theta_3$ and $\theta_4$), while A-B ($\theta_1$, $\theta_2$) and A-A ($\theta_5$) exchange interactions presents the lowest values, suggesting weakening of A-B and A-A exchange interactions. For x=0.10 all bond angles have values similar to x=0.00. This result suggests that, beyond x=0.06, we have a decreasing on the B-B interaction and strengthening of the A-B and A-A interactions.

### C. Optical Band Gap

The optical band-gaps of the FeO·Fe$_{(2-x)}$Nd$_x$O$_3$ nanoparticles were obtained by UV-Vis spectrophotometry at the temperature of 25ºC. The wavelength range of 350 nm to 850 nm using DH-2000-BAL (Mikropack) equipped with deuterium and tungsten halogen lamps. Attached to a DH-2000-BAL, an Ocean Optics® spectrometer USB4000 was used to measure the UV-Vis spectra [42-44]. The band gap of the nanoparticles is related to the optical gap ($E_g$) and photon energy (hυ) according to Eq. (12) [18, 48-50]:

$$[\alpha h\upsilon]^n = C(h\upsilon - E_g) \tag{12}$$

where C is a constant, and α is the linear absorption coefficient. The linear absorption coefficient α was calculated from the absorbance measurement A(hυ) as a function of the photon energy using Eq. (13) [48-50]:

$$\alpha(h\upsilon) = [A(h\upsilon) - A_{Scatt}(h\upsilon)]ln(10)/L \quad (13)$$

where $A_{scatt}(h\upsilon)$ is related to the Rayleigh-scattering contribution to the extinction measured data. $A_{scatt}(h\upsilon)$ was estimated as having a $\lambda^{-4}$ dependence, more details can be found elsewhere [42,44].

The value of *n* will be given according to the type of the electronic transition responsible for the absorption: for allowed indirect transition is *n* = 1/2; *n* = 3/2 for forbidden indirect transition; for allowed direct transition *n* = 2; *n* = 3 for forbidden direct transition [18,48-50]. The optical gap for both direct and indirect allowed transitions were studied in this work.

Here, A(hυ) is the experimental absorbance measured for each sample. The absorbance was measured using the UV-Vis spectrophotometry, and L = 1cm is the width of the cuvette cell. The optical band-gap was obtained by extrapolating the linear region of the Tauc plot (plot of $[\alpha h\upsilon]^n$ versus hυ) to a value of hυ= 0 [18, 48-50]. Tauc plot obtained for FF-REF and $Nd^{+3}$ substituted samples FF-ND1, FF-ND2, FF-ND3, FF-ND5 are shown in FIG. 2.

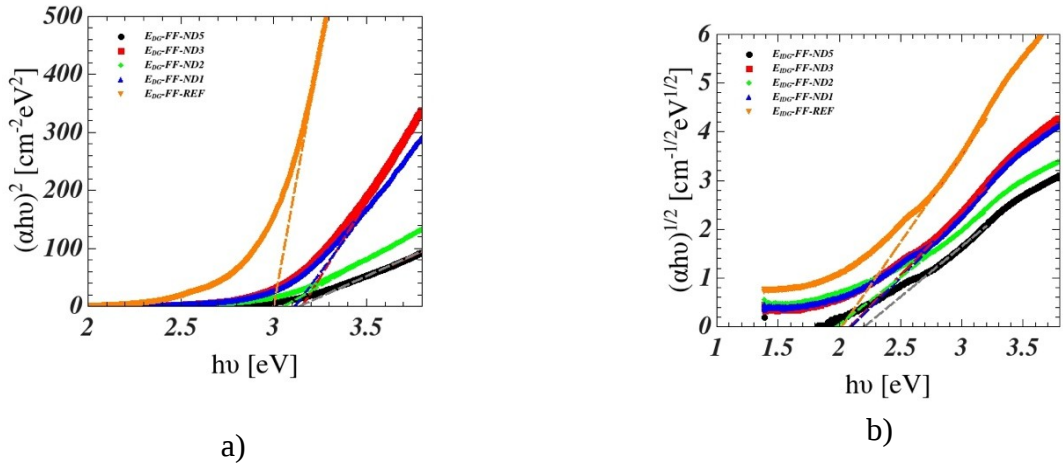

FIG. 2. (a) Tauc plots for $[\alpha h\upsilon]^2$ and (b) $[\alpha h\upsilon]^{1/2}$ versus the photon energy for determination of direct and indirect optical gaps, respectively.

Table V
Bandgaps of magnetic nanoparticles studied.

| Sample | Comp. (x)Nd | $E_d$ (eV) | $E_i$ (eV) |
|---|---|---|---|
| FF-REF | 0.00 | 3.00 (0.01) | 2.03 (0.01) |
| FF-ND1 | 0.02 | 3.12 (0.01) | 2.09 (0.01) |
| FF-ND2 | 0.04 | 3.05 (0.01) | 1.98 (0.01) |
| FF-ND3 | 0.06 | 3.16 (0.01) | 2.10 (0.01) |
| FF-ND5 | 0.10 | 3.13 (0.01) | 2.20 (0.01) |

Fontijn et al reported [51-53] that, for magnetite electronic transitions, $Fe^{2+}$ $[t_{2g}]$ → $Fe^{3+}$ $[e]$ occurring in the tetrahedral sites, the bandgap is 3.11 eV, while $Fe^{2+}$ $[t_{2g}]$ → $Fe^{3+}$ $[e_g]$ transitions occurring in the octahedral sites the bandgap is 1.94 eV. Their results were obtained through magneto-optical polar Kerr measurements for bulk single-crystalline $Fe_3O_4$ and $Mg^{2+}$ or $Al^{2+}$ substituted $Fe_3O_4$. These values are consistent with those obtained here. One can see in Table V, that for all samples with x ≠ 0.00, both $E_d$ and $E_i$ increase as x increases. We also point out that $E_d$ reaches the maximum value for x=0.06 (FF-ND3) rather than $E_i$ that reaches the maximum value for x=0.10.

J. Anghel and co-workers showed that exists a strong correlation between modifications in the lattice parameters and the bandgap energy for $Zn_{(1-x)}M_xO$ (M = Cr, Mn, Fe, Co, or Ni) samples [54]. In their studies they observed that the unit cell volume (obtained by XRD) and bandgap (obtained by spectrophotometry) reached the highest values of $Fe^{3+}$ substitution [54]. For the series of metals M that they studied, iron was the one with larger ionic radius, that leads to a higher lattice parameter and hence, unit cell volume. A similar trend was also observed elsewhere [55,56]. From XRD we found that the lattice parameter increases with increasing Nd-content and hence, unit cell volume, bond lengths ($d_{AL}$, $d_{BL}$) and hopping lengths, $L_A$ and $L_B$ between the magnetic ions at A-site and B-site. In this way, the increasing in the bandgap observed here, may be understood since the bandgap is directly proportional to the interatomic separation, although it is also possible that new electronic states may exist due to the presence of the dopant [55].

E. Transmission electronic microscopy (TEM)

Typical TEM micrographs of the samples investigated are shown in FIG. 3 and 4 for FF-REF and FF-ND3 samples, respectively. Typical TEM micrographs for FF-ND1, FF-ND2 and FF-ND5 samples is given in FIG. (S2), (S3) and (S4) (ESI), respectively. These results reveal a polydisperse size and shape distribution, as expected for co-precipitation synthesis method, with a broad number weighted-size distribution. The size distribution ranges from 5 to 30 nm and follows the log-normal distribution. The mean-number weighted diameters are given in Table VI. The polydispersity index (PDI = $\sigma(D_{TEM})/\langle D_{TEM} \rangle$) also were evaluated and they are given in the Table VI.

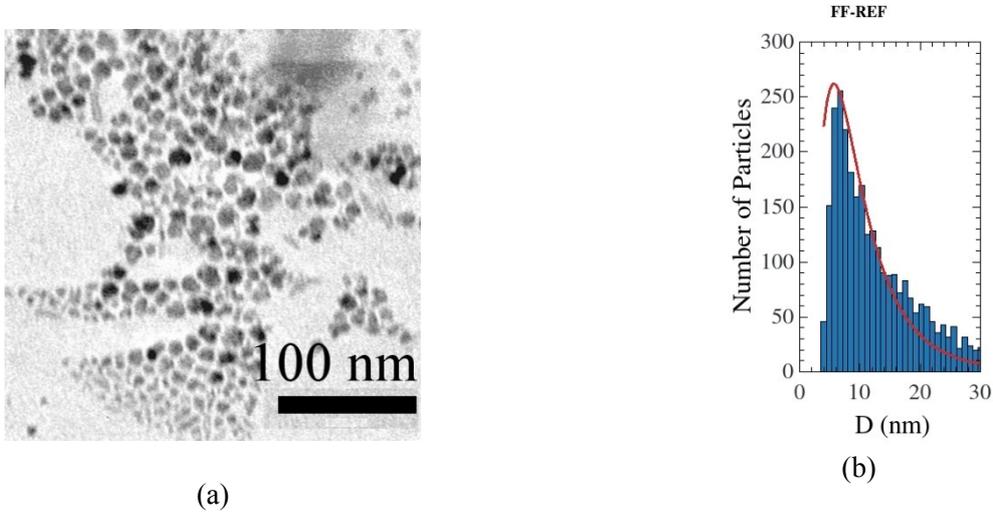

(a)                                  (b)

FIG. 3. (a) Typical TEM micrography and (b) Number-weighted diameter distribution of the FF-REF sample. The solid line is the log-normal fitting of size distribution given by Eq. (1).

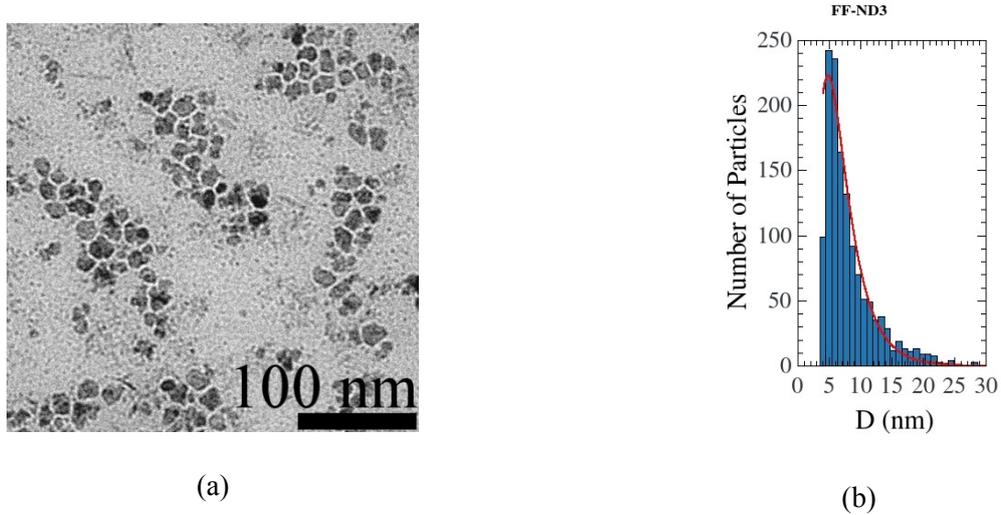

(a)                                  (b)

FIG. 4 (a) Typical TEM micrography and (b) Number-weighted diameter distribution of FF-ND3 sample. The solid line is the log-normal fitting of size distribution given by Eq. (1).

**Table VI**

Particles' mean diameter (*D*), distribution width (σ) and polydispersity and index (PDI) of the magnetic fluids investigated.

| Sample  | diameter *D* (σ) [nm] | PDI  |
|---------|-----------------------|------|
| FF-REF  | 12.81 (9.15)          | 0.71 |
| FF-ND1  | 12.50 (3.55)          | 0.28 |
| FF-ND2  | 11.67 (3.53)          | 0.30 |
| FF-ND3  | 7.99 (3.67)           | 0.46 |
| FF-ND5 | 9.06 (2.83) | 0.31 |

**F. Small Angle X-ray Scattering (SAXS)**

The direct obtained experimental scattering pattern by Small Angle X-ray Scaterring (SAXS) can provide several characteristic properties of the sample irradiated by the X-ray beam [2, 9]. For polydisperse systems, the experimental scattering pattern corresponds to an average that is performed over the particles present in solution [2, 9].

In the analysis of the scattered intensity data, it was assumed that $I(q)$ was produced by a system composed of globular nanoparticles. Therefore, $I(q)$ has a contribution from the particles' form-factor and the interaction between them (structure-factor). Assuming a model of polydisperse hard spheres with radius $r$, the scattered intensity is given by Eq. (14) [12]:

$$I_s(q) = Sc \int f_V(r)[V(r)]I_{sph}(q,r)dr \tag{14}$$

In this model, the scattering elements are considered an ensemble of polydisperse non-interacting spheres. The quantities r and V are the radius and volume for each sphere from the ensemble. The $f_v(r)$ is the normalized-volume weighted-radius distribution function. $I_{sph}$ is the normalized scattering intensity owing to a sphere of radius r and, Sc stands for a scaling factor [12]. The Gnom software was used to analyze the scattering intensity Is versus the scattering vector q (modulus of q) [57,12]. This approach was used to determine the volume-weighted size-distribution function $f_v(r)$ from the adjustment of the experimental scattering intensity data to $I_s(q)$, given by Eq. (14). The solid lines in FIG. 5 and 6 for FF-REF and FF-ND3, represent the best fit of Eq. (14) to the experimental data. For the remaining samples, namely FF-ND1, FF-ND2 and FF-ND5 the experimental data and the best fit are shown in FIG. (S5),(S6) and (S7)(ESI), respectively. The volume-weighted mean particle's diameter $\langle D_V \rangle$ was calculated according to Eq. (15) [12]:

$$\langle D_V \rangle = \frac{2 \int r f_V(r) dr}{\int f_V(r) dr} \tag{15}$$

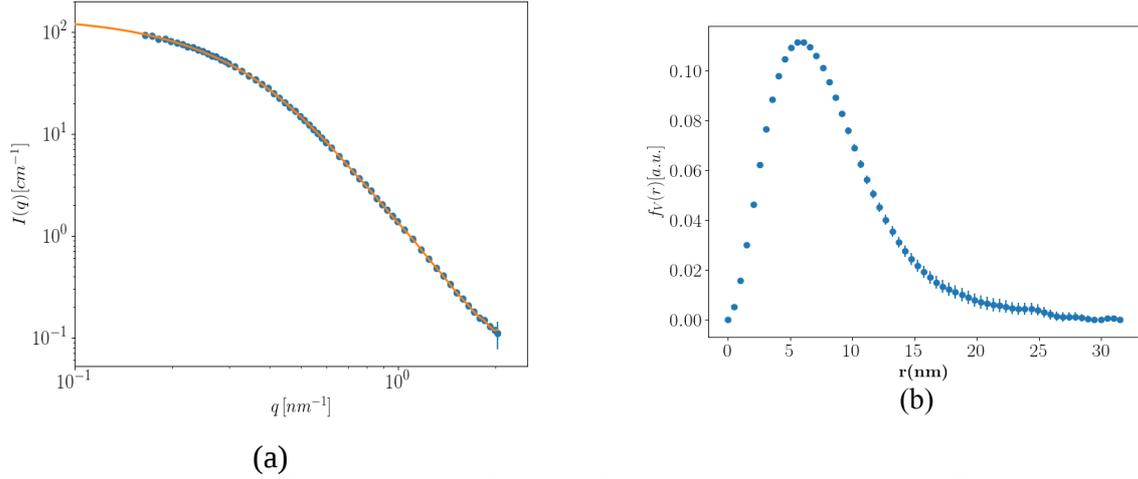

(a)

(b)

FIG. 5: (a) X-ray scattering intensity versus the scattering vector q (modulus). Solid line corresponds to best fit with Eq. (15); (b) Particle's radius numerical distribution function of sample FF-REF.

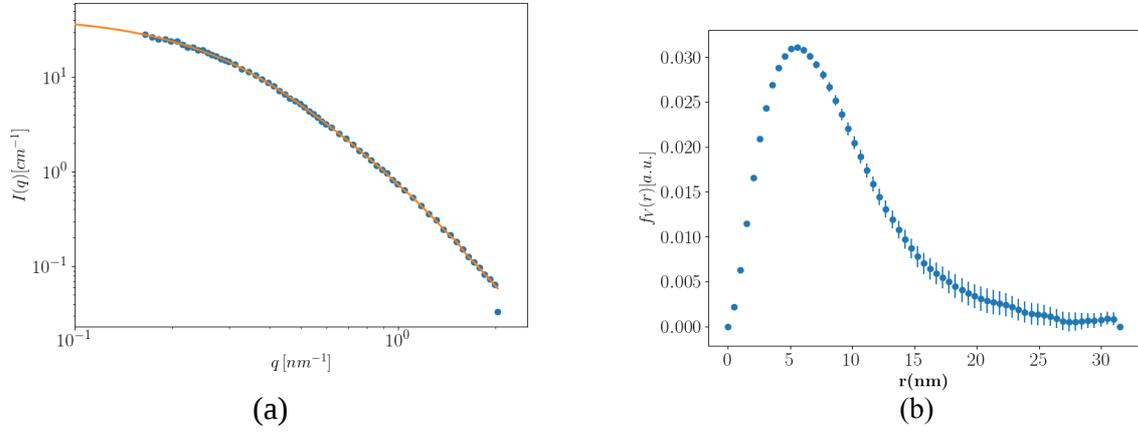

(a)

(b)

FIG. 6: (a) X-ray scattering intensity versus the scattering vector q (modulus). Solid line corresponds to best fit with Eq. (15); (b) Particle's radius numerical distribution function of sample FF-ND3.

Table VII

Nanoparticles' Mean diameter $\langle D_V \rangle$ determined by SAXS and width of the log-normal volume distribution function ($\sigma_V = 3\, \sigma_D$).

| Sample | Comp. (x)Nd | $\langle D_V \rangle$ (nm) | $\sigma_V$ |
|---|---|---|---|
| FF-REF | 0.00 | 18.48 | 1.83 |
| FF-ND1 | 0.02 | 28.74 | 1.86 |
| FF-ND2 | 0.04 | 29.15 | 1.83 |
| FF-ND3 | 0.06 | 20.32 | 2.07 |
| FF-ND5 | 0.10 | 28.84 | 1.77 |

The results given in Table VII for the volume-weighted average diameter measured by SAXS are consistent with those determined by XRD.

**G. Magnetization as a function of the applied field.**

FIG. 7 (a) and (b) show the measured magnetization (M) versus the applied field (H) for FeO·Fe$_{(2-x)}$Nd$_x$O$_3$ samples, at 6 K and 300 K, respectively. From these curves, the saturation magnetization (Ms), coercivity field (Hc) and remanence magnetization (Mr) are obtained and are given in Table VIII. The saturation magnetization increases in a steady manner upon increasing the Nd$^{3+}$ ion molar ratio from 0.00 up to 0.06, reaching the maximum value of 105.8±0.4 Am$^2$/Kg at x= 0.06. It's worth to mention that the result for saturation magnetization value are higher than that of the bulk material. In this case, suggesting the existence of a highly ordered spin configuration distributed across the volume of the magnetic nanoparticles.

The M(H) curves for magnetic nanoparticles in superparamagnetic (SPM) regime (FIG. 7-(b)) are fitted with a Langevin function (Eq. (16)), weighted with the particle-size distribution function (PDF) [58,59]:

$$M(H,T) = N_{SC} \int_0^\infty \frac{x k_B T}{\mu_0 H} \left[ \left( coth(x) - \frac{1}{x} \right) + cH \right] PDF(D_V) dD_V \qquad (16)$$

where M(H,T) is the magnetization of the magnetic nanoparticles at the temperature T (300 K), N$_{sc}$ is a scaling constant, M$_S$ represents the saturation magnetization of the magnetic nanoparticles and $x = \mu_0 M_s V_V H / k_B T$ [59]. The quantity V$_V$, stands for the volume-weighted average volume of the nanoparticles given by $V_V = \frac{\pi}{6} \langle D_V \rangle^3$ [52]. Here we used, $\langle D_V \rangle$ from the volume-weighted average radius and hence diameter, measured by SAXS [12].

In the model expressed by Eq. (16), the magnetization M as a function of field H has contributions from superparamagnetic (SPM) and paramagnetic (PM) particles [31]. The quantity *c* is the paramagnetic contribution (linear with the magnetic field, H [36]). We found that the SPM regime contributes with 95-97% for all samples, while only 3-5% comes from the PM regime [36].

The MNPs are expected to have a nonmagnetic layer around the magnetic core (core-shell model) [60, 61]. The thickness of this magnetically inert shell was evaluated from the volume-weighted size distribution function, according to Eq. (17) [60, 61]:

$$M_S = M_{S_{bulk}} \left( 1 - \frac{6\delta}{\langle D_V \rangle} \right) \qquad (17)$$

where δ is the thickness of the shell, M$_S$ is saturation magnetization of the nanoparticles

$M_{Sbulk}$ stands for the saturation magnetization of the bulk material [60-62] and the reciprocal of the volume-weighted average diameter ($1/\langle D_V \rangle$) of the particles [60,61]. From Eq. (16) we found that the thickness of the magnetically inert layer of the materials studied here, ranges from 0.5 Å to 7.8 Å.

The Neel's theory for collinear ferro/ferrimagnetism of two sub-lattices model, predicts the net magnetic moment per formula unit (f.u.) according to Eq. (17):

$$n_{Neel} = M_{octa} - M_{tetra} \tag{18}$$

where $M_{octa}$ and $M_{tetra}$ are the magnetic moments of B (octahedral) and A (tetrahedral) sites in $\mu_B$ units (Bohr magneton) [36, 31]. Based on the site occupancy obtained from the XRD cation distribution and magnetic moment of 5, 4 and 3.2 $\mu_B$ [63] for $Fe^{3+}$, $Fe^{2+}$ and $Nd^{3+}$ ions, respectively, the $n_{Neel}$ values were calculated using Eq. (18) and are given in Table VIII. On the other hand, the magnetic moment per formula unit in Bohr magneton unit $n_{exp}$ (experimental) can be calculated from the saturation magnetization Ms according to Eq. (18):

$$n_{exp} = \frac{M_S \cdot MW}{5585} \tag{19}$$

where MW is the molecular weight. The calculated $n_{Neel}$ magneton number using the XRD data agrees with the experimentally obtained magneton number from the M-H loops. The results are comparable to those determined by Eq. (19). The calculated values of magneton number $n_{Neel}$ and $n_{exp}$ are comparable. Furthermore, the values of $n_{Neel}$ and $n_{exp}$ show same trend, that is, increases with x=0.00 to x =0.04, reaches the maximum value at x = 0.06 and decreases for x =0.10.

The Yafet-Kittel angles (αY-K) described by equation Eq. (20) can give us insight about the type of magnetic ordering for $FeO \cdot Fe_{(2-x)}Nd_xO_3$ samples [30, 36]. The quantity $n_{exp}$ is given by Eq. (19) while $M_{tetra}$ and $M_{octa}$ are given by Eq. (18). The Yafet-Kittel angles (αY-K) obtained were 21.6 °, 16.3 °, 14.1 °, 0.0 ° and 0.0 ° for x=0.00, 0.02, 0.04, 0.06 and 0.10, respectively. From x=0.00 to x=0.04 αY-K decreases with increasing $Nd^{3+}$ substitution. The decrease in αY-K for this samples is due to a noncollinear type of magnetic ordering, since the Yafet-Kittel angles are different from zero. The existence of non-zero αY-K suggest a model of canted-spin magnetization, which should have a triangular spin arrangement and is suitable on the B-site that leads to a reduction in the A–B exchange interaction and enhancement in the B-B exchange interaction [30]. For x=0.06, there is a transition to the Néel-type magnetic ordering. This result suggests a reinforcement of a dominant A-B super-exchange interaction and these results are corroborated by those found in the inter-ionic bond angles of Table IV.

$$\alpha_{YK} = cos^{-1}\left(\frac{n_{exp}+M_{tetra}}{M_{octa}}\right) \tag{20}$$

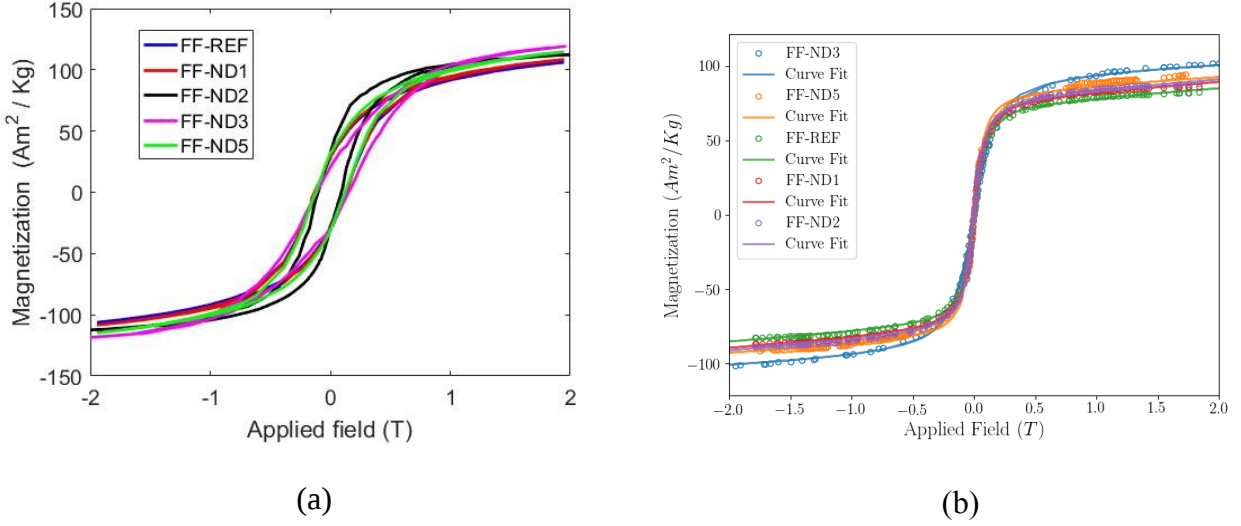

(a)          (b)

FIG. 7. Magnetization curves of FF-REF, FF-ND1, FF-ND2, FF-ND3 and FF-ND5 nanoparticles a) at 6 K and b) at 300 K, the dots are experimental results and solid lines are their size-weighted Langevin fits.

**Table VIII**
Magnetic parameters extracted from M(H) curves.

| Sample | Comp. (x) | Saturation magnetization ($M_s$) Am²/Kg | Remnant magnetization ($M_r$) Am²/Kg | Coercivity ($H_c$) mT | Squareness ratio (S) | Magnetic anisotropy energy density ($K_{eff}$) $10^3$ J/m³ | Bohr magneton number ($\mu_B$) exp. | Neel's Bohr magneton number ($\mu_B$) XRD. |
|---|---|---|---|---|---|---|---|---|
| FF-REF | 0.00 | 81.9 (0.8) | 30.4 (0.2) | 156 | 0.37 | 33.1 | 3.40 (0.01) | 4.00 (0.44) |
| FF-ND1 | 0.02 | 86.1 (0.3) | 28.3 (0.1) | 156 | 0.33 | 35.1 | 3.60 (0.01) | 3.96 (0.44) |
| FF-ND2 | 0.04 | 87.7 (0.3) | 38.6 (0.2) | 121 | 0.44 | 27.9 | 3.69 (0.01) | 3.92 (0.43) |
| FF-ND3 | 0.06 | 105.8 (0.4) | 24.3 (0.1) | 157 | 0.23 | 44.8 | 4.49 (0.02) | 4.08 (0.45) |
| FF-ND5 | 0.10 | 91.0 (0.3) | 32.8 (0.2) | 156 | 0.36 | 40.2 | 3.97 (0.02) | 3.95 (0.42) |

For magnetite ($Fe_3O_4$), below the Verwey Temperature ($T_V$), the magnetocrystalline anisotropy is expected to be uniaxial [62, 64]. Therefore, the experimental coercivity field and saturation magnetization are related to the effective anisotropy constant $K_{eff}$ through Eq. (21) [62, 64]:

$$K_{eff} = \frac{H_C M_s}{2} \tag{21}$$

To determine the squareness ratio (S), Eq. (22) was employed [65, 14]:

$$S = \frac{M_r}{M_s} \tag{22}$$

The evaluated values of the magnetocrystalline anisotropy constant ($K_{eff}$) and S are given in Table VIII. For all samples S < 0.5, which indicates uniaxial anisotropy contribution in the ~~prepared~~ $FeO \cdot Fe_{(2-x)}Nd_xO_3$ nanoparticles [60, 62-65]. The overall values of $K_{eff}$ obtained are of the same order of magnitude of that from the magnetite bulk (1.1 – 1.3 $10^4$ J/m$^3$) [62-65] however, we have obtained values 3 to 4 times higher.

### H. Zero-field cooled and field-cooled (ZFC-FC)

FIG. 8(a)-(e) show the temperature dependence of the magnetization in low external applied field, the zero-field cooled curve (ZFC) and field-cooled curve (FC) for FF-REF, FF-ND1, FF-ND2, FF-ND3 and FF-ND5, respectively. In the ZFC-FC protocol the MNPs were frozen in the absence of the magnetic field, fast enough that the random orientation of their easy axis is preserved [12, 13]. The system was superparamagnetic (SPM) at room temperature, and the magnetization curves $M(T)$ were collected in ZFC and FC modes. The overall shape of the ZFC-FC curves indicate weak interaction between particles. In the ambit of the non-interacting particles' model, the blocking-temperature distribution function $f(T_B)$ is expected to be broad [12]. One can use the Stoner-Wohlfarth model to describe uniaxial, single-domain [8] and non-interacting particles to obtain the blocking-temperature distribution function given by Eq. (23) [12]. FIG. 8(f)-(i) show $f(T_B)$ for FF-REF, FF-ND1, FF-ND2, FF-ND3 and FF-ND5, respectively. A log-normal distribution function was used to fit these results and obtain the mean blocking temperature $\langle T_B \rangle$ and standard deviation $\sigma_T$ [12]. For FF-REF, FF-ND1, FF-ND2, FF-ND3 and FF-ND5 samples, the average blocking temperatures were 40.2 K, 28.8 K, 23.6, 25.1 and 41.7 K, respectively, at an applied field of 5 mT. Note that FIG. 8(a)-(e) also show that the maximum of ZFC curve for all cases is close to the temperature ~~in~~ which the ZFC-FC curves split. This effect is due to particle dipole-dipole interaction.

$$f(T_B) \propto \frac{-d[M_{FC} - M_{ZFC}]}{dT} \tag{23}$$

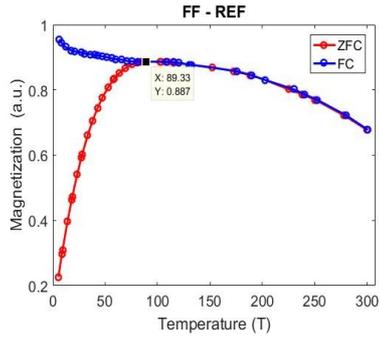

(a)

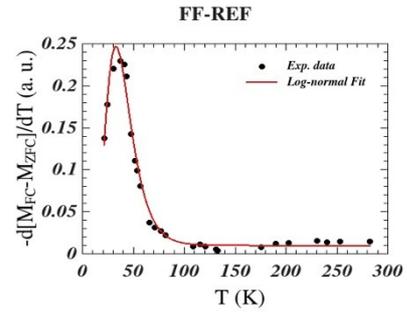

(f)

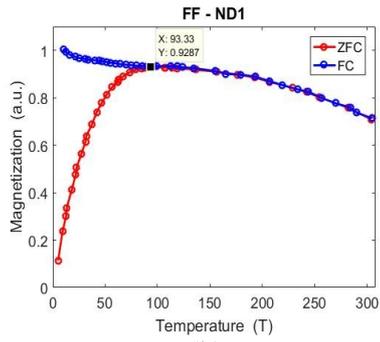

(b)

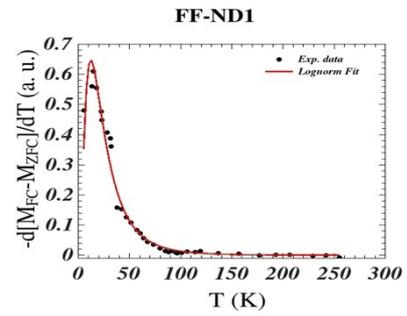

(g)

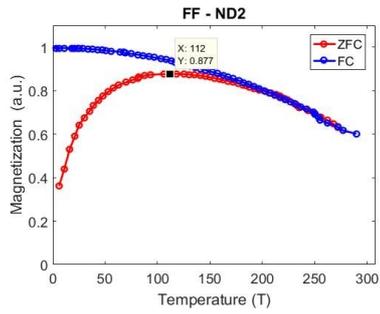

(c)

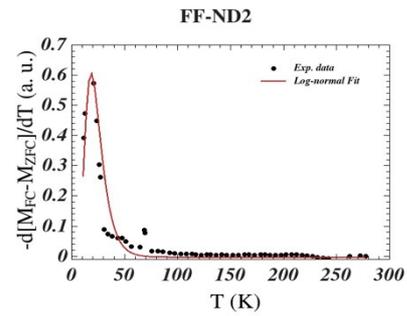

(h)

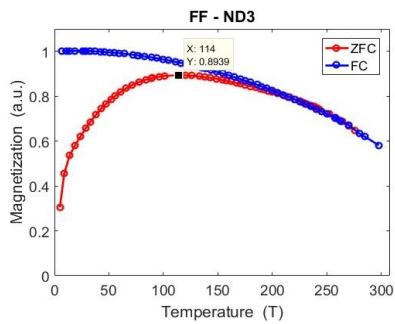

(d)

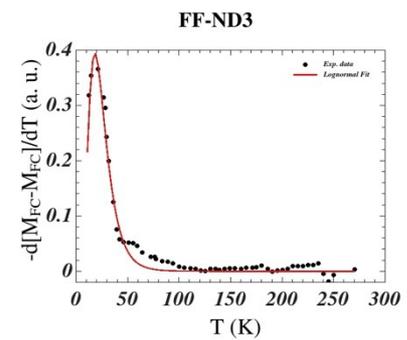

(i)

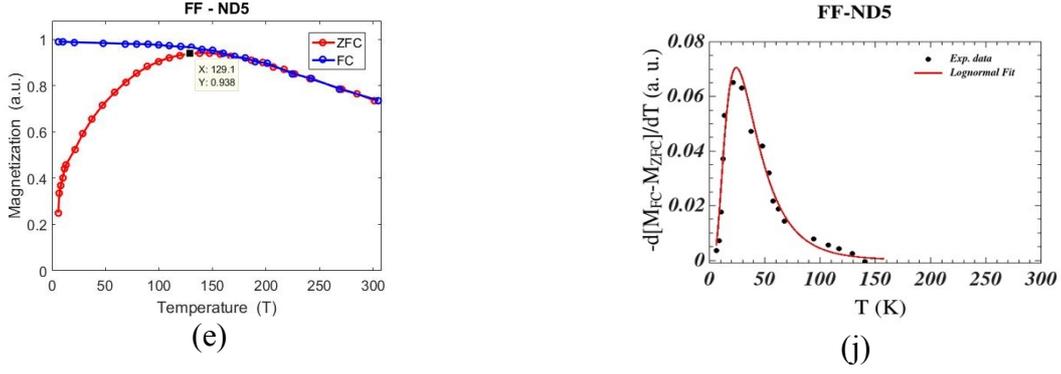

(e)　　　　　　　　　　　　　　　　　　(j)

FIG. 8. (a, b, c, d and e) ZFC-FC magnetization curves, where (●) blue circles represent the FC susceptibility, whereas red circles (●) symbols represent the ZFC susceptibility for FF-REF, FF-ND1, FF-ND2, FF-ND3, and FF-ND5, respectively. 8 (f, g, h, i and j) Temperature derivative [-d($M_{FC}$-$M_{ZFC}$)/dT] of the difference between FC and ZFC magnetization curves for samples FF-ND, FF-RC and FF-REF, respectively.

The log-normal width $\sigma_T$, obtained from the fitting to the experimental data – FIG. 8(f)-(i) – were used to evaluate the number of interacting particles ($N_i$) and the correlation volume ($\Lambda^3$), defined according to Eqs. (24) and (25) [12, 13]:

$$N_i = \left(\frac{\sigma_V}{\sigma_T}\right)^2 \tag{24}$$

$$\Lambda^3 = \frac{N_i \langle V \rangle}{\phi} \tag{25}$$

where $\sigma_V$, $\sigma_T$ are the width of the volume and blocking temperature distribution functions. The quantity $\langle V_V \rangle$ is the average volume defined as:

$$\langle V_V \rangle = \frac{\pi}{6} \langle D_V \rangle^3 \tag{26}$$

where $\langle D \rangle$ is the volume-weighted average diameter measured by SAXS. It's worth to mention, that according with M. El-Hilo [58], the volume-weighted diameter distribution $f_v(D)$ should converts to volume-weighted distribution $f_v(V)$ with $\sigma_V = 3\sigma_D$ and average volume-weighted by volume given by Eq.(26). The quantity $\phi$ is the volumetric fraction of particles in the solution [8, 12, 13], ~ 0.3% for all the samples investigated.

### Table IX
Width of volume and blocking temperature distribution $\sigma_V$, $\sigma_T$. Number of interacting particles $N_i$. The correlation volume ($\Lambda^3$) and correlation length ($\Lambda$).

| Sample | $\sigma_V$ | $\sigma_T$ | $N_i$ | $\Lambda^3$ ($10^{-23}$ m$^3$) | $\Lambda$ (nm) |
|---|---|---|---|---|---|
| FF-REF | 1.83 | 0.37 | 5 | 5.4 | 176 |
| FF-ND1 | 1.86 | 0.76 | 3 | 10.1 | 216 |

| FF-ND2 | 1.83 | 0.41 | 5 | 19.3 | 226 |
| FF-ND3 | 2.07 | 0.46 | 5 | 6.6 | 187 |
| FF-ND5 | 1.77 | 0.60 | 3 | 12.4 | 231 |

Table IX, shows estimation of the number of interacting particles ($N_i$) inside the correlation volume ($\Lambda^3$) and correlation length ($\Lambda$) [12]. The results did not show correlation between $N_i$ and the size of the particles. However, the correlation volume ($\Lambda^3$) and correlation length ($\Lambda$) are smaller for FF-REF and FF-ND3, which are those with smaller volume-weighted average diameter measured by SAXS. The correlation volume ($\Lambda^3$) and correlation length ($\Lambda$) also are larger for those particles with larger volume-weighted average diameter (FF-ND1, FF-ND2 and FF-ND5). This is due to the fact that for larger particles, the interparticle distance increases, as the mean-particle diameter increases [12]. Note that, the correlation length ($\Lambda$) is out of the range accessible in our SAXS experiments, that is, about ~60 nm.

## V. Summary and Conclusions

Magnetic fluid based on magnetite were synthesized and XRD patterns of the samples investigated revealed a single-phase cubic spinel structure of Fd3m space group. From the cation distribution, the theoretical lattice parameter ($a_{th}$), tetrahedral bond length ($d_{AL}$), octahedral bond length ($d_{BL}$), tetrahedral edge ($d_{AE}$), octahedral edge ($d_{BE}$), unshared octahedral edge ($d_{BEU}$) increase with the increase in Nd-content. Since the $Fe^{3+}$ (0.65Å) is smaller than $Nd^{3+}$ (0.98 Å), the replacement of $Fe^{3+}$ leads to an increase in rA and rB. Moreover, $L_A$ and $L_B$ increase with higher Nd-content. The results showed that $Nd^{3+}$ ions are placed in both sites at different concentrations. Therefore, it was possible to observe modifications in structural parameters like bond lengths; shared and unshared edges. The increasing in the bandgap may be interpreted as a result of the higher interatomic separation with the doping. TEM micrographs reveal a polydisperse size and shape distribution of particles, as expected for co-precipitation method, with a broad number-weighted size distribution. The results for volume-weighted average diameter measured by SAXS are consistent with those determined by XRD. From the M-H Loops we found that the SPM regime contributes with 95-97% for all samples, while only 3-5% contribution comes from the PM regime. The saturation magnetization increases in a steady manner upon increasing the $Nd^{3+}$ ion molar ratio from 0.00 up to 0.06, reaching the maximum value of 105.8±0.4 $Am^2/Kg$ at x= 0.06. It's worth to mention that the result for saturation magnetization value are higher than that of the bulk material. In this case, suggesting the existence of a highly ordered spin configuration distributed across the volume of the magnetic nanoparticles. The squareness values for all samples are less than 0.5, which indicates uniaxial-anisotropy contribution in the $FeO·Fe_{(2-x)}Nd_xO_3$ nanoparticles. The overall values of $K_{eff}$ obtained for all samples studied, have the same order of magnitude of that from the bulk magnetite ($1.1 - 1.3 \cdot 10^4$ J/m$^3$) however, we have obtained values 3 to 4 times higher. The magnetic measurements indicate the existence of low interaction between the MNPs at the concentrations investigated.

**ACKNOWLEGMENTS**

This research was financial supported by Brazil's agencies INCT/CNPq (Conselho Nacional de Desenvolvimento Científico e Tecnológico; Grant Number: 465259/2014-6), INCT/FAPESP (Fundação de Amparo à Pesquisa do Estado de São Paulo; Grant Number: 14/50983-3), INCT/CAPES (Coordenação de Aperfeiçoamento de Pessoal de Nível Superior; Grant Number: 88887.136373/2017-00), and INCT-Fcx (Instituto Nacional de Ciência e Tecnologia de Fluidos Complexos).

# Electronic supplementary information.

## Magnetic, Structural and cation distribution studies on FeO·Fe$_{(2-x)}$Nd$_x$O$_3$ (x=0.00, 0.02, 0.04, 0.06 and 0.1) nanoparticles.

W. W. R. Araujo[*], C. L. P. Oliveira, G. E. S. Brito, A. M. Figueiredo Neto

Institute of Physics, University of São Paulo, Rua do Matão 1371, 05508-090, São Paulo, SP, Brazil.

J. F. D. F. Araujo,

Department of Physics, Pontifical Catholic University of Rio de Janeiro, RJ, Brazil.

The mean ionic radii in the tetrahedral (rA) and octahedral (rB) sites were calculated by using Eqs. (S1) and (S2) [1-4]:

$$rA = C_{ANd} r_{Nd^{+3}} + C_{AFe} r_{Fe^{+3}} \quad (S1)$$

$$rB = \frac{1}{2}\left[ C_{BFe} r_{Fe^{+2}} + C_{BNd} r_{Nd^{+3}} + C_{BFe} r_{Fe^{+3}} \right] \quad (S2)$$

where, $C_{Ai}$ ( i=Fe$^{3+}$,Nd$^{3+}$ ) are the concentration of ions in the tetrahedral site and $C_{Bi}$ ( i= Fe$^{2+}$,Fe$^{3+}$,Nd$^{3+}$ ) the concentration of ions in the octahedral site. The values of the ionic radius for $r_{Nd3+}$, $r_{Fe3+}$ and $r_{Fe3+}$ were taken from the literature, 0.98 Å, 0.67 Å, and 0.49 Å, respectively.

The value of the oxygen positional parameter $u$ can be determined with Eq. (S3) [1-4]:

$$u = \left| \frac{(rA+R_O)}{a\sqrt{3}} + \frac{1}{4} \right| \quad (S3)$$

where $R_O$ is the radius of oxygen ion (1.32 Å).

The theoretical lattice constant ($a_{th}$) is calculated by using Eq. (S4):

$$a_{th} = \frac{8}{3\sqrt{3}}\left[ (rA+R_O) + \sqrt{3}(rB+R_O) \right] \quad (S4)$$

The values of tetrahedral bond length (d$_{AL}$), octahedral bond length (d$_{BL}$), tetrahedral edge length (d$_{AE}$), shared (d$_{BE}$) and unshared (d$_{BEU}$) octahedral edge length, the tetrahedral and octahedral jump length (L$_A$ and L$_B$ ) are determined using Eqs. (S5)-(S11) [1-4]. The results are given in Table III.

$$d_{AL} = a\sqrt{3}\left(u - \frac{1}{4}\right) \quad (S5)$$

$$d_{BL} = a\sqrt{3u^2 - \frac{11u}{4} + \frac{43}{64}} \quad (S6)$$

$$d_{AE} = a\sqrt{2}\left(2u - \frac{1}{2}\right) \tag{S7}$$

$$d_{BE} = a\sqrt{2}(1 - 2u) \tag{S8}$$

$$d_{BEU} = a\sqrt{\left(4u^2 - 3u + \frac{11}{16}\right)} \tag{S9}$$

$$L_A = a\frac{\sqrt{3}}{4} \tag{S10}$$

$$L_B = a\frac{\sqrt{2}}{4} \tag{S11}$$

The configuration of ion pairs in the spinel structure, as well as their distances and angles has been illustrated in FIG. (S1) [1-4].

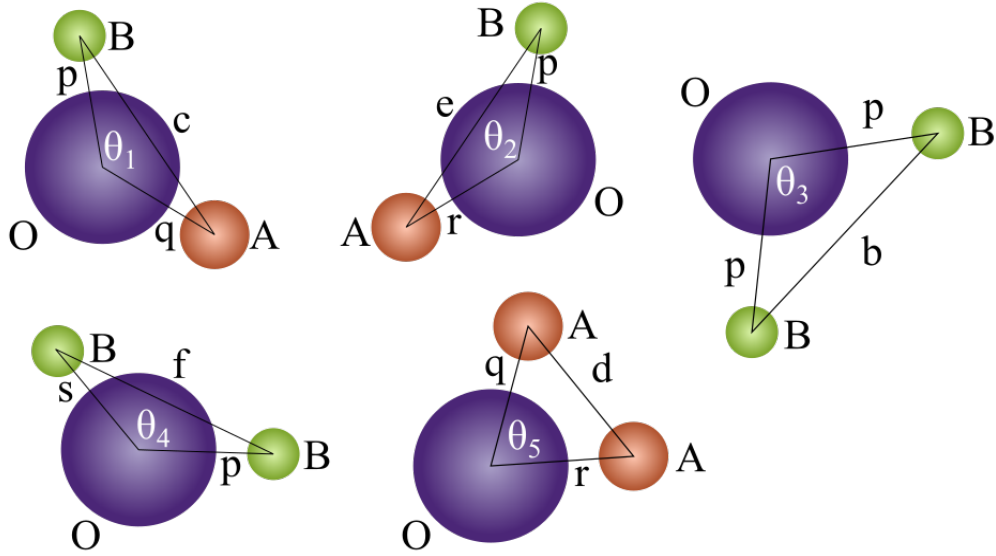

FIG. S1. Configuration of the ion pairs in spinel ferrites with favorable distances and angles.

The bond lengths values for Me-Me (metal-metal) and Me-O (metal-oxygen) were calculated theoreticaly using the following sets of equations, Eqs. (S12)-(S20) [3]:

Me-Me:

$$b = \frac{a\sqrt{2}}{4} \tag{S12}$$

$$c = \frac{a\sqrt{11}}{8} \tag{S13}$$

$$d = \frac{a\sqrt{3}}{4} \tag{S14}$$

$$e = \frac{3a\sqrt{3}}{8} \tag{S15}$$

$$f = \frac{a\sqrt{6}}{4} \tag{S16}$$

Me-O:

$$p = a\sqrt{\left(\frac{1}{16} - \frac{\delta}{2} + 3\delta^2\right)} \tag{S17}$$

$$q = a\sqrt{3}\left(\frac{1}{8} + \delta\right) \tag{S18}$$

$$r = a\sqrt{3}\left(\frac{1}{4} + \delta\right) \tag{S19}$$

$$s = a\sqrt{\left(\frac{1}{16} + \frac{\delta}{2} + 3\delta^2\right)} \tag{S20}$$

The values of bond angles were calculated using Eq. (S21)-(S25), as follows:

$$\theta_1 = \cos^{-1}\left(\frac{p^2 + q^2 - c^2}{2pq}\right) \tag{S21}$$

$$\theta_2 = \cos^{-1}\left(\frac{p^2 + r^2 - e^2}{2pr}\right) \tag{S22}$$

$$\theta_3 = \cos^{-1}\left(\frac{p^2 - b^2}{2p^2}\right) \tag{S23}$$

$$\theta_4 = \cos^{-1}\left(\frac{p^2 + s^2 - f^2}{2ps}\right) \tag{S24}$$

$$\theta_5 = \cos^{-1}\left(\frac{r^2 + q^2 - d^2}{2rq}\right) \tag{S25}$$

where b, c, d, e, f, p, q, r and s represent the parameters of ion pair distances associated with angles $\theta_1$, $\theta_2$, $\theta_3$, $\theta_4$, and $\theta_5$ for spinel ferrites structures [1-4]. The parameter $\delta$ is the deviation of the oxygen positional parameter u from ideal position $u_{ideal}$ = (3/8, 3/8, 3/8) i.e. $\delta = u - u_{ideal}$ [2].

Typical TEM micrographs for FF-ND1, FF-ND2 and FF-ND5 samples is given in FIG. (S2), (S3) and (S4), respectively.

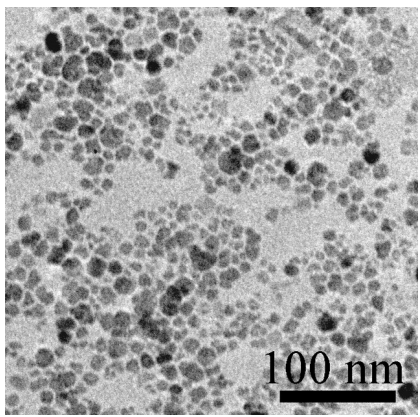
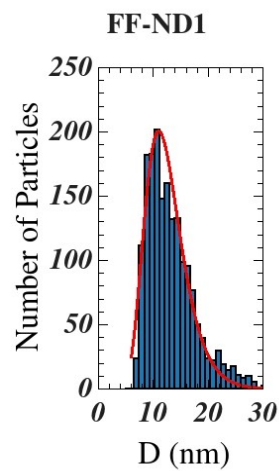

(a)

(b)

FIG. S2. (a) Typical TEM micrography and (b) Number-weighted diameter distribution for FF-ND1 samples. The solid line is the log-normal fitting of size distribution given by Eq. (1).

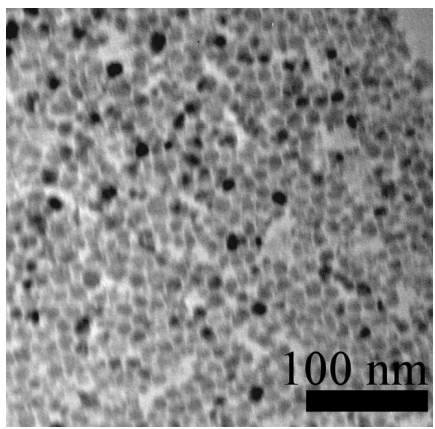
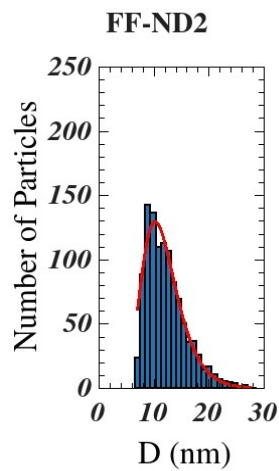

(a)

(b)

FIG. S3 (a) Typical TEM micrography and (b) Number-weighted diameter distribution for FF-ND2 samples. The solid line is the log-normal fitting of size distribution given by Eq. (1).

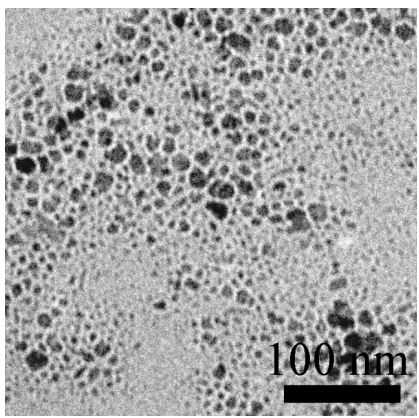
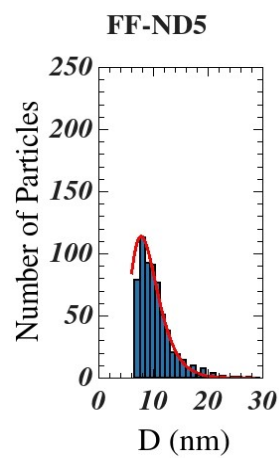

(a)

FIG. S4 (a) Typical TEM micrography and (b) Number-weighted diameter distribution for FF-ND5 samples. The solid line is the log-normal fitting of size distribution given by Eq. (1).

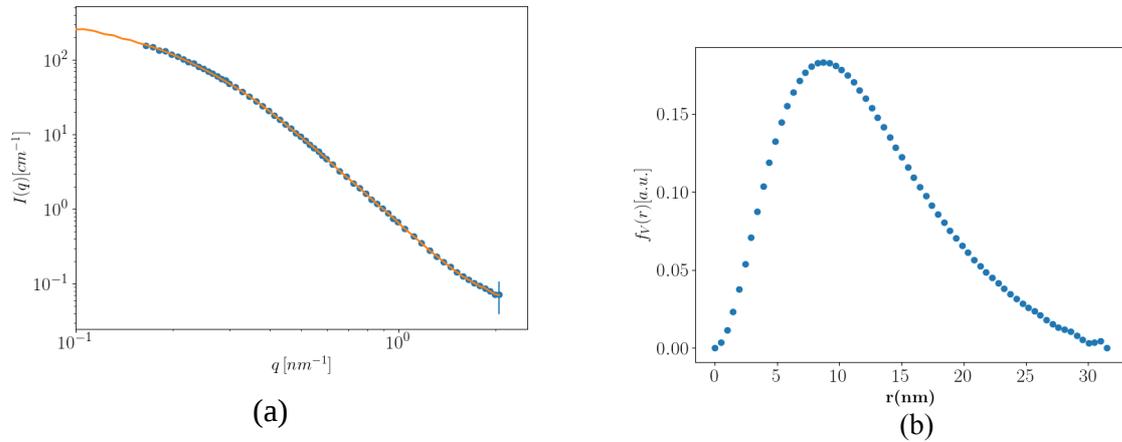

(a) (b)

FIG. S5: (a) X-ray scattering intensity versus the scattering vector q (modulus). Solid line corresponds to best fit with Eq. (15); (b) Particle's radius numerical distribution function of sample FF-ND1.

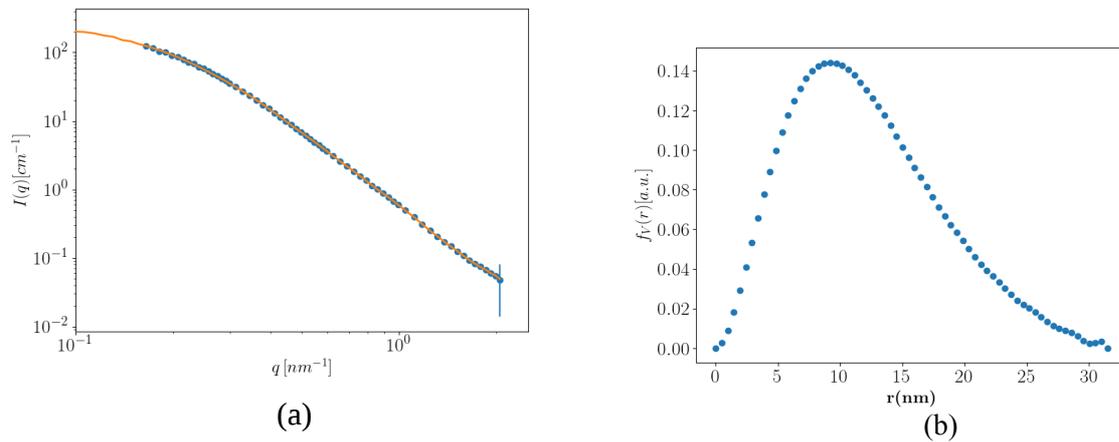

(a) (b)

FIG. S6: (a) X-ray scattering intensity versus the scattering vector q (modulus). Solid line corresponds to best fit with Eq. (15); (b) Particle's radius numerical distribution function of sample FF-ND2.

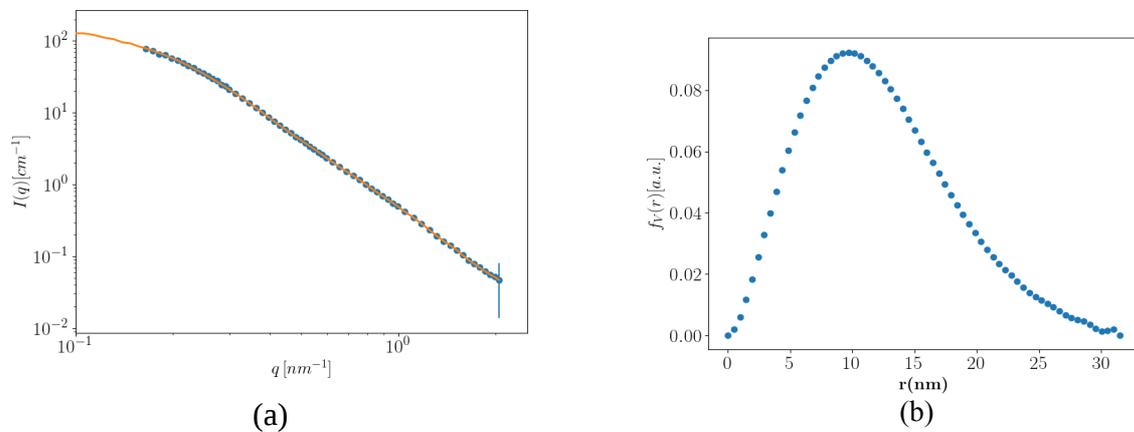

(a) (b)

FIG. S7: (a) X-ray scattering intensity versus the scattering vector q (modulus). Solid line corresponds to best fit with Eq. (15); (b) Particle's radius numerical distribution function of sample FF-ND5.